\documentclass[aps,prb,superscriptaddress,twocolumn,floatfix]{revtex4-1}
\usepackage{graphicx}
\usepackage{amssymb}
\usepackage{amsfonts}
\usepackage{amsmath}

\begin{document}

\title{Superconductivity under pressure in the Dirac semimetal PdTe$_2$}

\author{H. Leng} \email{h.leng@uva.nl}\affiliation{Van der Waals - Zeeman Institute, University of Amsterdam, Science Park 904, 1098 XH Amsterdam, The Netherlands}
\author{A. Ohmura} \affiliation{Pacific Rim Solar Fuel System Research Center, Niigata University, 8050, Ikarashi 2-no-cho, Nishi-ku, Niigata, 950-2181, Japan}
\affiliation{Faculty of Science, Niigata University, 8050, Ikarashi 2-no-cho, Nishi-ku, Niigata, 950-2181, Japan}
\author{L. N. Anh} \affiliation{International Training Institute for Materials Science, Hanoi University of Science and Technology, 1 Dai Co Viet Road, Ha Noi, Vietnam}
\author{F. Ishikawa} \affiliation{Faculty of Science, Niigata University, 8050, Ikarashi 2-no-cho, Nishi-ku, Niigata, 950-2181, Japan}
\author{T. Naka} \affiliation{National Institute for Materials Science, Sengen 1-2-1, Tsukuba, Ibaraki 305-0047, Japan}
\author{Y. K. Huang} \affiliation{Van der Waals - Zeeman Institute, University of Amsterdam, Science Park 904, 1098 XH Amsterdam, The Netherlands}
\author{A. de Visser} \email{a.devisser@uva.nl} \affiliation{Van der Waals - Zeeman Institute, University of Amsterdam, Science Park 904, 1098 XH Amsterdam, The Netherlands}

\date{\today}

\begin{abstract}
The Dirac semimetal PdTe$_2$ was recently reported to be a type-I superconductor ($T_c = $1.64~K, $\mu_0 H_c (0) = 13.6$~mT) with unusual superconductivity of the surface sheath. We here report a high-pressure study, $p \leq 2.5$~GPa, of the superconducting phase diagram extracted from ac-susceptibility and transport measurements on single crystalline samples. $T_c (p)$ shows a pronounced non-monotonous variation with a maximum $T_c = $1.91~K around 0.91~GPa, followed by a gradual decrease to 1.27~K at 2.5~GPa. The critical field of bulk superconductivity in the limit $T \rightarrow 0$, $H_c(0,p)$, follows a similar trend and consequently the $H_c(T,p)$-curves under pressure collapse on a single curve: $H_c(T,p)=H_c(0,p)[1-(T/T_c(p))^2]$. Surface superconductivity is robust under pressure as demonstrated by the large superconducting screening signal that persists for applied dc-fields $H_a > H_c$. Surprisingly, for $p \geq 1.41$~GPa the superconducting transition temperature at the surface $T_c^S$ is larger than $T_c$ of the bulk. Therefore surface superconductivity may possibly have a non-trivial nature and is connected to the topological surface states detected by ARPES. We compare the measured pressure variation of $T_c$ with recent results from band structure calculations and discuss the importance of a Van Hove singularity.

\end{abstract}

\maketitle
%\begin{multicols}{2}

\section{Introduction}
The family of layered transition metal dichalcogenides attracts much attention, because of the wide diversity of fascinating electronic properties. One of the present-day research interests is the possibility to realize novel quantum states as a result of the topological non-trivial nature of the electronic band structure~\cite{Soluyanov2015,Huang2016,Yan2017,Bahramy2018}. Especially, it has been proposed that these materials host a generic coexistence of type-I and type-II three dimensional Dirac fermion states~\cite{Bahramy2018}. An interesting example in this respect is PdTe$_2$ that has been classified as a type-II Dirac semimetal following a concerted examination of \textit{ab-initio} electronic structure calculations and angle-resolved photoemission spectroscopy (ARPES) experiments~\cite{Liu2015a,Fei2017,Noh2017,Bahramy2018,Clark2017}. In a type-II Dirac semimetal the Hamiltonian breaks Lorentz invariance because the energy dispersion relations, \textit{i.e.} the Dirac cone, are tilted~\cite{Soluyanov2015}. The Dirac point is then the touching point between the electron and hole pockets and a nearly flat band may form near the Fermi level. Moreover, PdTe$_2$ is a superconductor below 1.6~K~\cite{Guggenheim1961,Leng2017}, which solicits the intriguing question whether superconductivity is promoted by the nearly flat band and consequently has a topological nature~\cite{Fei2017}. Topological non-trivial superconductors attract much interest since it is predicted these may host protected Majorana zero modes at the surface (for recent reviews see Refs.~\onlinecite{Sato&Fujimori2016,Sato&Ando2017}). This in turn offers a unique design route to make devices for topological quantum computation.

Superconductivity in PdTe$_2$ was discovered in 1961~(Ref.~\onlinecite{Guggenheim1961}), but was not investigated in detail until 2017, when Leng \textit{et al.}~\cite{Leng2017} carried out comprehensive magnetic and transport experiments on single-crystals. Unexpectedly, dc-magnetization measurements, $M(H)$, revealed that PdTe$_2$ is a bulk type-I superconductor, which was further embodied by the observation of the differential paramagnetic effect in the ac-susceptibility measured in applied  magnetic dc-fields. The critical field $H_c(T)$ follows the standard quadratic temperature variation with $\mu_0 H_c(0) = 13.6$~mT. The possibility of type-I superconductivity in Dirac materials was recently investigated by Shapiro \textit{et al}.~\cite{Shapiro2018} employing a microscopic pairing theory for an arbitrary tilt parameter of the Dirac cone. For PdTe$_2$ these authors concluded type-I superconductivity is feasible for a tilt parameter $k=2$. Another interesting aspect of PdTe$_2$ is the observation of surface superconductivity, as evidenced by large screening currents in the ac-susceptibility for applied dc-fields $H_a > H_c$~\cite{Leng2017}. The critical field for surface superconductivity $H_c ^S$ does not follow the standard Saint-James - de Gennes expression $H_{c3} = 2.39 \times \kappa H_c$~\cite{Saint-James&deGennes1963}, where $\kappa$ is the Ginzburg-Landau parameter. This led to the proposal~\cite{Leng2017} that superconductivity of the surface sheath might have a topological nature and originates from topological surface states detected by ARPES~\cite{Liu2015a,Noh2017}. More recently, specific heat~\cite{Amit2018} and magnetic penetration depth~\cite{Salis2018,Teknowijoyo2018}, measurements have been conducted. These confirm conventional weak-coupling Bardeen-Cooper-Schrieffer (BCS) superconductivity, with a full gap in the bulk. At the same time zero-field scanning tunneling microscopy (STM) and spectroscopy (STS) experiments~\cite{Das2018,Clark2017} lend further support for the absence of in-gap states, which seems to rule out topological superconductivity at the surface. Dominant $s$-wave superconductivity was also concluded from tunneling spectroscopy experiments on side junctions~\cite{Voerman2019}. Nonetheless, the uncommon type-I behavior for a binary compound, and the unexplained superconductivity of the surface sheath, justify a further in-depth examination of the superconducting properties of PdTe$_2$.

We here report the results of a high-pressure investigation of the superconducting phase diagram of PdTe$_2$ single crystals ($p \leq 2.5$~GPa). Combined resistivity and ac-susceptibility measurements show $T_c$ increases at low pressures, then passes  through a maximum of 1.91~K around 0.91~GPa, and subsequently decreases at higher pressure. The critical field for $T \rightarrow 0$, $H_c(0,p)$, follows a similar behavior and consequently the $H_c(T)$-curves at different pressures collapse on a single curve. Under pressure superconductivity maintains its type-I character. Surface superconductivity is robust under pressure as demonstrated by the large superconducting screening signal that persists for applied dc-fields $H_a > H_c$. Surprisingly, for $p \geq 1.41$~GPa the superconducting transition temperature of the surface, $T_c^S$, is larger than $T_c$ of the bulk. Therefore surface superconductivity may possibly have a non-trivial nature and is related to the topological surface states detected by ARPES\cite{Liu2015a,Noh2017,Clark2017}. The initial increase of $T_c$ with pressure is at variance with the smooth depression predicted by recent electronic structure calculations~\cite{Xiao2017}.

\section{Experiment}
\begin{figure}[t]
\centering
\includegraphics[width=8cm]{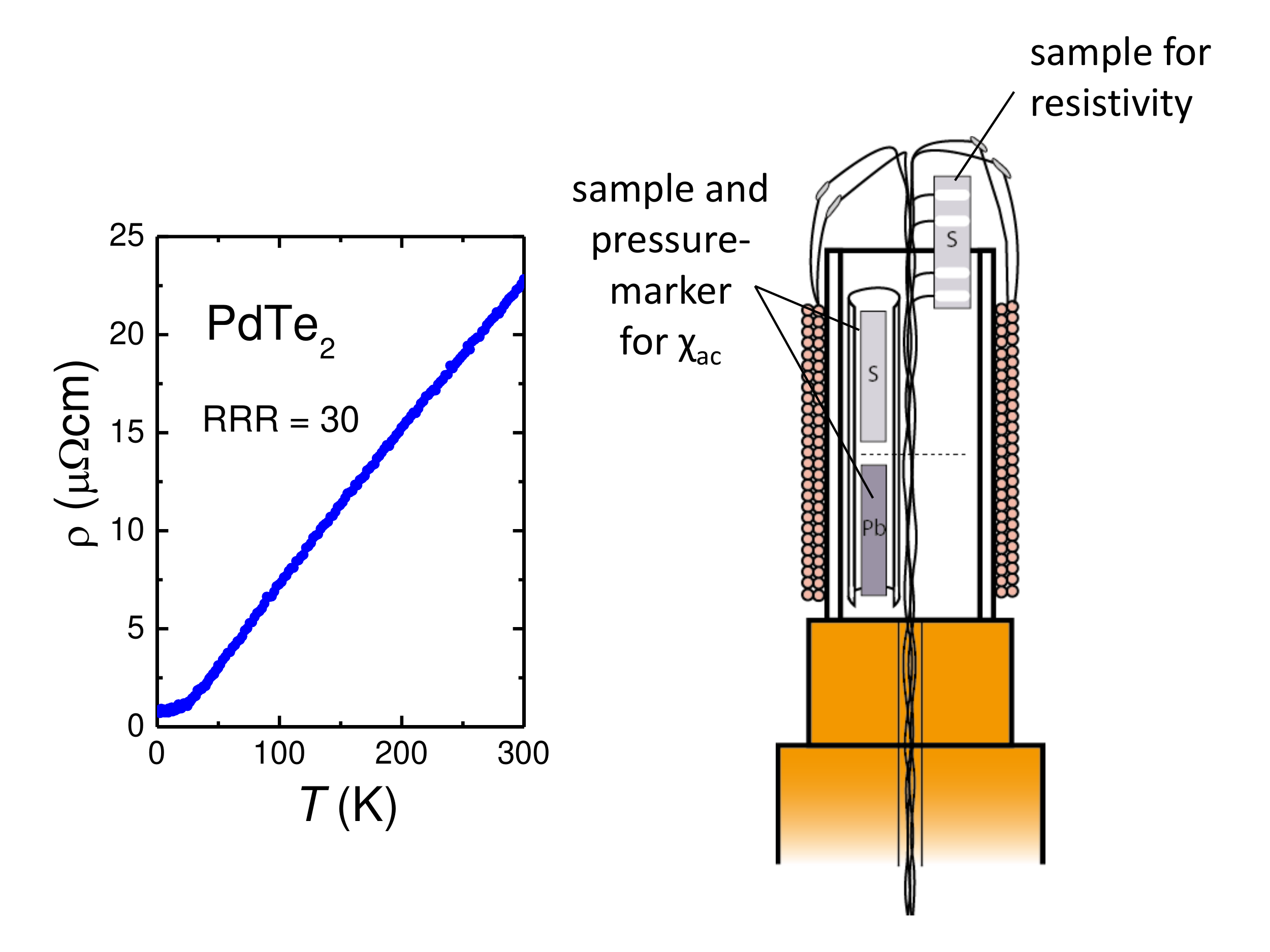}
\caption{Left: Resistivity of PdTe$_2$ measured with the current in the basal plane. Right: Pressure plug with samples and $\chi_{ac}$ coils mounted (schematic).}
\label{fig:figure1}
\end{figure}

The crystals used for our high pressure study were taken from the single-crystalline boule prepared by the modified Bridgman technique~\cite{Lyons1976} and characterized in Ref. \onlinecite{Leng2017}. Powder X-ray diffraction confirmed the trigonal CdI$_2$ structure (spacegroup $P\bar{3}m1$~\cite{Thomassen1929}. Scanning electron microscopy (SEM) with energy dispersive x-ray (EDX) spectroscopy showed the proper 1:2 stoichiometry within the experimental resolution of 0.5\%. Laue backscattering was used to orient the crystals. Standard four-point resistance measurements were performed in a Physical Property Measurement System (PPMS, Quantum Design) at temperatures down to 2~K. The resistivity, $\rho (T)$, of our crystals shows metallic behavior. A typical trace in the temperature range 2-300~K is shown in Fig.~\ref{fig:figure1}. The residual resistance ratio $R$(300K)/$R$(2K) = 30.

Electrical resistance, $R(T,H)$, and ac-susceptibility, $\chi_{ac}(T,H)$, measurements under high-pressure were performed utilizing a clamp-type piston-cylinder cell, which has a double-layer made of Cu-Be and NiCrAl alloys. The single crystal sizes for $R(T,H)$ and $\chi_{ac}(T,H)$ were $\sim 2.3 \times 1.0 \times 0.18$~mm$^3$ and $\sim 2.9 \times 1.0 \times 0.67$~mm$^3$, respectively. Both samples were mounted on a plug and loaded into a Teflon capsule together with coils and a pressure-transmitting medium, Daphne oil 7373, for hydrostatic compression. A schematic drawing of the plug with samples and coil is shown in Fig.~\ref{fig:figure1}. The generated pressure in the capsule relating to each load was estimated from the calibration data for this cell, which was obtained from the pressure variations of superconducting transition temperatures of lead and tin in previous experiments~\cite{Slooten2009,Bay2012}. We carried out the compression experiments on the crystals twice, first up to a pressure of 1.24~GPa and in a second run up to 2.49~GPa.

Typical experimental conditions are as follows. The high-pressure cell was compressed at room temperature and then cooled down to about 0.3~K using a $^3$He refrigerator (Oxford Instruments Heliox VL). Electrical resistivity was measured by a resistance bridge (Linear Research LR-700) using a low-frequency ac method with an excitation current $I = 300 \mu$A. In order to investigate the field-suppression of $T_c$, a magnetic field was applied along the current, parallel to the $a$-axis. For ac-susceptibility measurements, a small cylinder, composed of an excitation coil and a pick-up coil in which the sample is situated, was prepared. The in-phase and out-of-phase signals were detected in the driving field  $\mu_0 H_{ac} = 0.0047$~mT with a frequency of $f_{ac} = 313$~Hz using a lock-in amplifier (EG\&G Instruments Model 7260). Measurements were made in zero field and in applied dc-fields using a superconducting magnet. Special care was taken to reduce the remnant field of the superconducting magnet to close to zero, since our PdTe$_2$ crystals show type-I superconductivity.

Overall the resistivity, $\rho(T)$, measured in the temperature range 2-300~K showed little variation with pressure and remained metallic. However, the absolute $\rho$-value at 300~K decreases smoothly with respect to pressure to about 80\% of the ambient pressure value at the highest pressure 2.49~GPa.

\section{Results}

\subsection{Pressure-temperature phase diagram}

\begin{figure}[t]
\centering
\includegraphics[width=8.5cm]{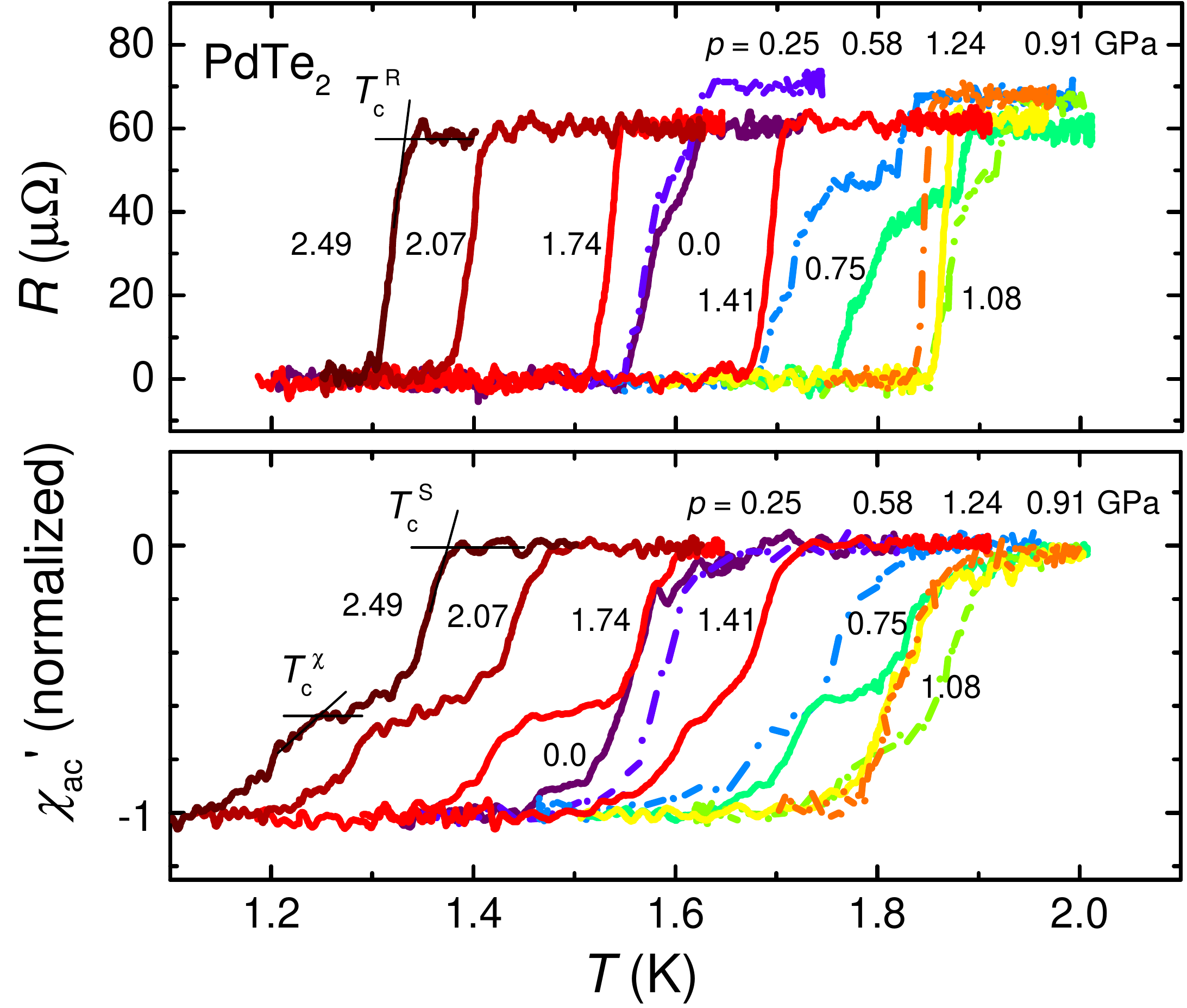}
\caption{Resistance and ac-susceptibility (normalized to $-1$ in the superconducting state) of single-crystalline PdTe$_2$ as a function of temperature around $T_c$ at pressures up to 2.49 GPa as indicated. The data were taken in two pressure runs (see text): dashed-dotted lines for the first run ($p$-values listed above the curves) and solid lines for the second run ($p$-values listed adjacent to the curves). The yellow curve in both panels was taken at 1.08~GPa. The curves at 0 GPa were measured after releasing the pressure in the second run. $T_c^R$ is determined from the onset of superconductivity in $R(T)$ as indicated for $p=2.49~$GPa by the thin solid lines. For $p \geq 1.41$~GPa the onset of diamagnetic screening in $\chi_{ac}(T)$ is attributed to surface superconductivity at $T_c^S$, and the further drop signals bulk superconductivity at $T_c^{\chi}$, as indicated for $p=2.49$~GPa. See text.}
\label{fig:figure2}
\end{figure}

\begin{figure}[b]
\centering
\includegraphics[width=7cm]{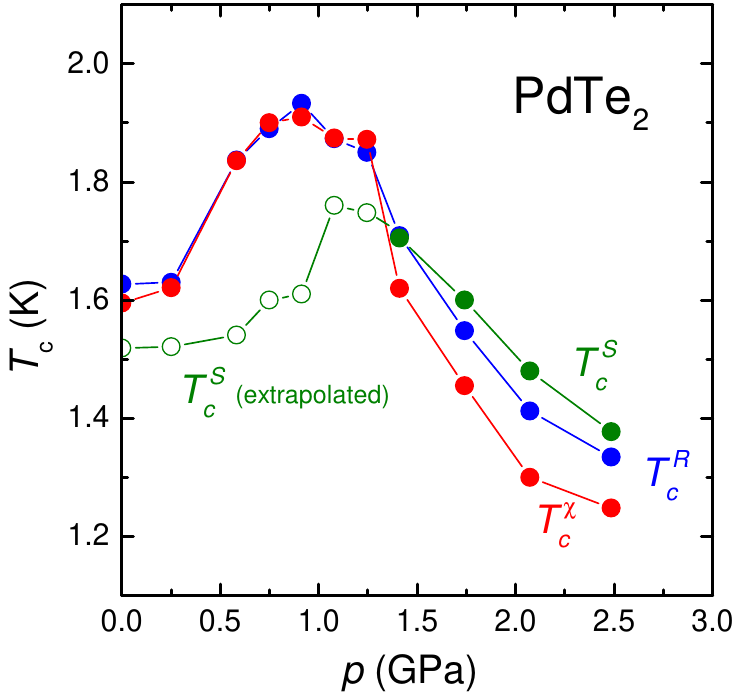}
\caption{Pressure variation of the superconducting transition temperature of PdTe$_2$ as determined from resistance, $T_c^R$ (blue symbols), and from ac-susceptibility, $T_c^{\chi}$ (red symbols). $T_c^S$ denotes surface superconductivity (open and closed green symbols). Open symbols are determined by extrapolation. Note that for $p \geq 1.41$~GPa $T_c^S > T_c^{bulk} = T_c^{\chi}$ (see text).}
\label{fig:figure3}
\end{figure}

The overall results of the two pressure runs are reported in Fig.~\ref{fig:figure2}. In the first run data were taken at pressures of 0.25, 0.58, 0.91 and 1.24~GPa. Here the normal state resistance $R_N \simeq 70~\mu \Omega$. For the second run new voltage contacts were made on the crystal resulting in $R_N  \simeq 60~\mu \Omega$. The applied pressures are 0.75, 1.08, 1.41, 1.74, 2.07 and 2.49~GPa. We remark the zero-pressure data were measured after releasing the pressure. Also, the value of the ac-susceptibility differed somewhat between different cool downs and between the two pressure runs. For clarity all the $\chi_{ac}$ data in the lower panel of Fig.~\ref{fig:figure2} are normalized to $-1$ in the superconducting state.

The resistance curves around $T_c$ at ambient pressure and $p=0.25$~GPa show a double structure which becomes more pronounced with increasing pressure. However, for $p \geq 1.08$~GPa the superconducting transition is sharp. We attribute the double structure in $R(T)$ at low pressures to parts of the crystal responding differently to pressure, because of an inhomogeneity, rather than to a pressure gradient. We remark that previous resistance experiments on crystals taken from the same single-crystalline boule revealed a single sharp superconducting transition at ambient pressure~\cite{Leng2017}. A similar behavior is observed in $\chi_{ac}(T)$ with relatively sharp, single transitions at pressures of 1.08 and 1.24~GPa. However, for $p \geq 1.41$~GPa the transition becomes structured again with an onset temperature of superconductivity larger than $T_c$ deduced from the resistivity curves (top panel). As we will demonstrate in the next Section, at these pressures the initial screening step is attributed to surface superconductivity~\cite{Leng2017}, while the ensuing second step with a full diamagnetic screening is attributed to bulk superconductivity.

The first important result is that superconductivity is enhanced under pressure with a maximum value $T_c =$~1.91~K around 0.91~GPa and a gradual depression of $T_c$ at higher pressures. This is illustrated in Fig.~\ref{fig:figure3}, where we trace $T_c^R (p)$ extracted from the resistance data. Here we use the onset transition temperatures determined by extrapolation of the $R(T)$-curves just below $T_c$ to the normal state plateau values, as shown for the 2.49~GPa curve in Fig.~\ref{fig:figure2}. The same analysis for the $\chi_{ac}(T)$-data shows $T_c^{\chi} (p)$ tracks $T_c^R (p)$ closely up to 1.24~GPa, see Fig.~\ref{fig:figure3}. However, for $p \geq 1.41$~GPa it is the second, lower in temperature, transition in $\chi_{ac} (T)$ that is attributed to bulk superconductivity and tracks $T_c^R (p)$. The agreement between $T_c^{\chi} (p)$ and $T_c^R (p)$ obtained with different techniques on two different crystals is good. Lastly, the temperature of surface superconductivity, $T_c^S (p)$, is traced in Fig.~\ref{fig:figure3}. For $p \leq 1.41$~GPa $T_c^S (p)$ is obtained from the field-temperature phase diagrams by extrapolating $T_c^S (H)$ to zero field, as reported in Ref.~\onlinecite{Leng2017} and presented in the following Section. For $p \geq 1.41$~GPa we take $T_c^S (p)$ from the onset of the upper transition in $\chi_{ac}(T)$. This tells us the transition temperatures for the bulk and surface have a distinct pressure variation, and for $p \geq 1.41$~GPa $T_c^S > T_c^{bulk} $. This underpins surface superconductivity in PdTe$_2$ is a unique, robust feature.

\begin{figure}[t]
\centering
\includegraphics[width=8.5cm]{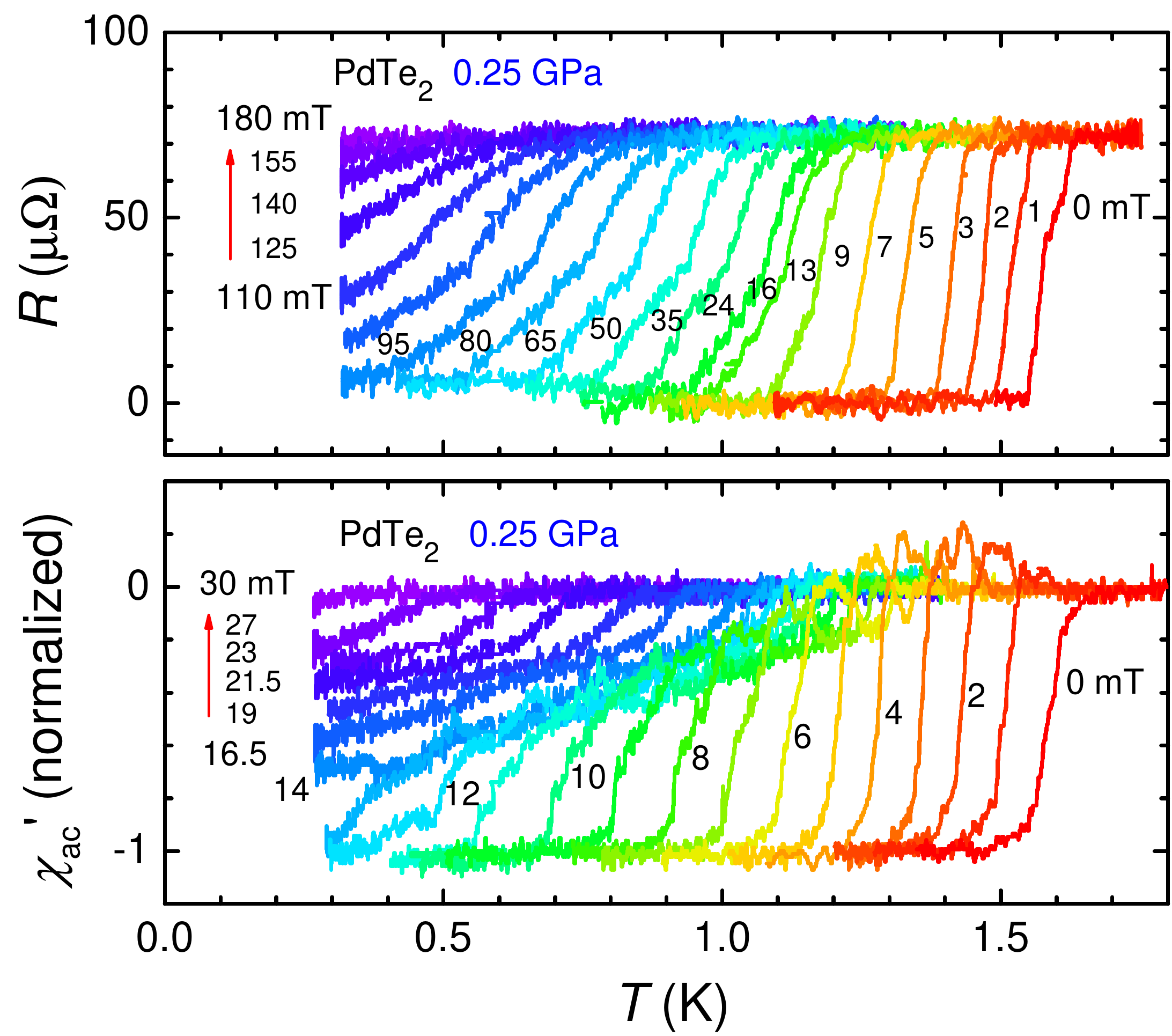}
\caption{Upper panel: Resistance of PdTe$_2$ as a function of temperature at a pressure $p = 0.25$~GPa measured in applied magnetic fields $\mu_0 H_a \parallel I \parallel a$. Curves from right to left are taken in fields of 0, 1, 2, 3, 5, 7, 9, 13, 16, 24, 35, 50, 65, 80, 95, 110, 125, 140, 155 and 180~mT. Lower panel: Ac-susceptibility at $p = 0.25$~GPa measured in applied magnetic fields. Curves from right to left in 0~mT to 14~mT with 1~mT steps and in 16.5, 19, 21.5, 23, 27 and 30~mT. }
\label{fig:figure4}
\end{figure}

\subsection{Field-temperature phase diagram}

In order to investigate the pressure dependence of the superconducting phase diagram in the $H$-$T$ plane we have measured at each pressure the resistance and ac-susceptibility in applied dc-fields, $H_a$. A typical data set taken at $p=0.25$~GPa is shown in Fig.~\ref{fig:figure4}. In the lower panel with $\chi_{ac}$-data the zero-field curve shows $T_c = 1.63$~K. In small applied fields a peak appears just below $T_c$ due to the differential paramagnetic effect (DPE). This peak signals the field induced intermediate state~\cite{Leng2017}. It shifts to lower temperatures with increasing field and for higher fields is progressively depressed because of an additional screening signal that precedes the DPE peak. The additional screening is attributed to superconductivity of the surface sheath~\cite{Leng2017}. Partial screening is still visible at 27~mT, but has nearly vanished at $\mu_0 H_a = 30$~mT down to 0.3~K. Consequently, in the limit $T \rightarrow 0$ $H_c^S(0) > H_c(0)$. In the upper panel, with $R(T)$ data, the transition is first rapidly depressed with field up to $\mu_0 H_a \approx 13$~mT, but then the depression rate decreases, the transition broadens and signals of superconductivity persist up to $\mu_0 H_a \approx 180$~mT. We remark this field is much larger than $H_c(0)$ or $H_c^S(0)$. The robustness of superconductivity in resistance measurements was also observed at ambient pressure, with a critical field, $H_c^R(0)$, equal to $\sim 0.3$~T~\cite{Leng2017}.

\begin{figure}[b]
\centering
\includegraphics[width=8.5cm]{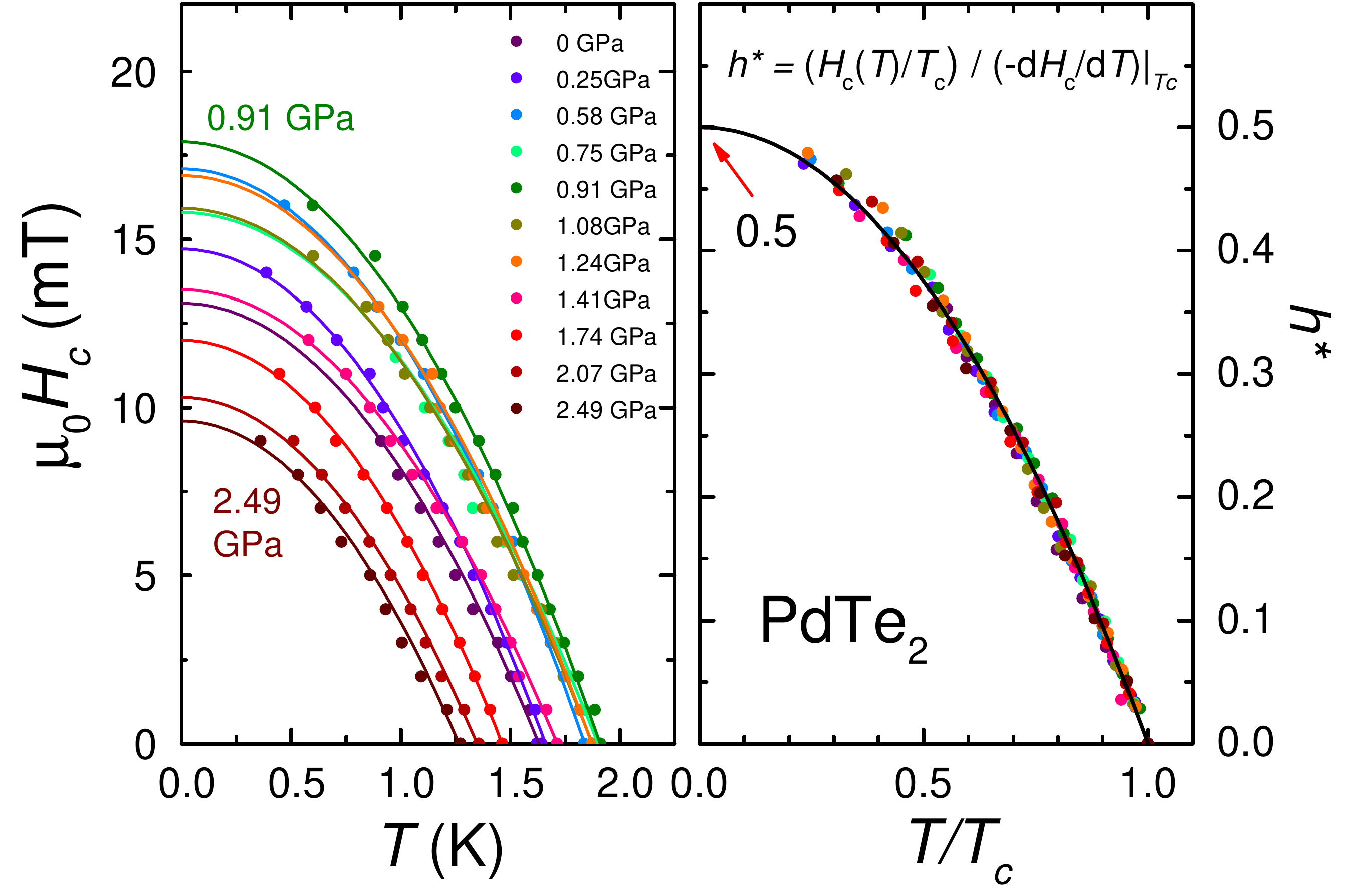}
\caption{Left panel: Critical field $H_c(T)$ for type-I superconductivity in PdTe$_2$ at pressures between 0 and 2.49~GPa as indicated. The solid lines represent $H_c(T) = H_c(0)[1-(T/T_c)^2]$ at different pressure, where $T_c = T_c^{\chi}$ is the bulk superconductivity transition temperature extracted from the $\chi_{ac}$-data in applied fields. Right panel: Reduced plot $h^* =(H_c(T)/T_c)/(-dH_c/dT)|_{T_c}$ \textit{versus} $T/T_c$. The solid line represents $h^* = 0.5 \times [1-t^2]$. See text.}
\label{fig:figure5}
\end{figure}

In the following paragraphs we present the $H$-$T$ phase diagrams determined from the $R(T)$- and $\chi_{ac}(T)$-data in applied fields, measured up to 2.49~GPa. The phase diagram at 0.25~GPa is extracted from Fig.~\ref{fig:figure4}. Additional data sets are presented in the Supplemental Material (SM)~\cite{SM}.

In Fig.~\ref{fig:figure5} we present the critical field for bulk superconductivity $H_c(T)$. The data are obtained by tracing the $T_c^{\chi}$-values as a function of the applied field. The solid lines in Fig.~\ref{fig:figure5} represent $H_c(T) = H_c(0)[1-(T/T_c)^2]$ at different pressures, where $T_c = T_c^{\chi}$. The quadratic temperature variation is consistent with type-I superconductivity. In fact all the data under pressure collapse on one single curve, $h^*(t)$, as shown in the right panel of Fig.~\ref{fig:figure5}. Here the standard expression for plotting $H_c(T)$ in a reduced form is applied, with $h^* =(H_c(T)/T_c)/(-dH_c/dT)|_{T_c}$ where $t=T/T_c$~\cite{Bay2012b}. For a type-I superconductor $h^*(0) = 0.5$. The collapsed curve $h^*(t)$ shows type-I superconductivity persists over the whole pressure range.

\begin{figure}[t]
\centering
\includegraphics[width=8.5cm]{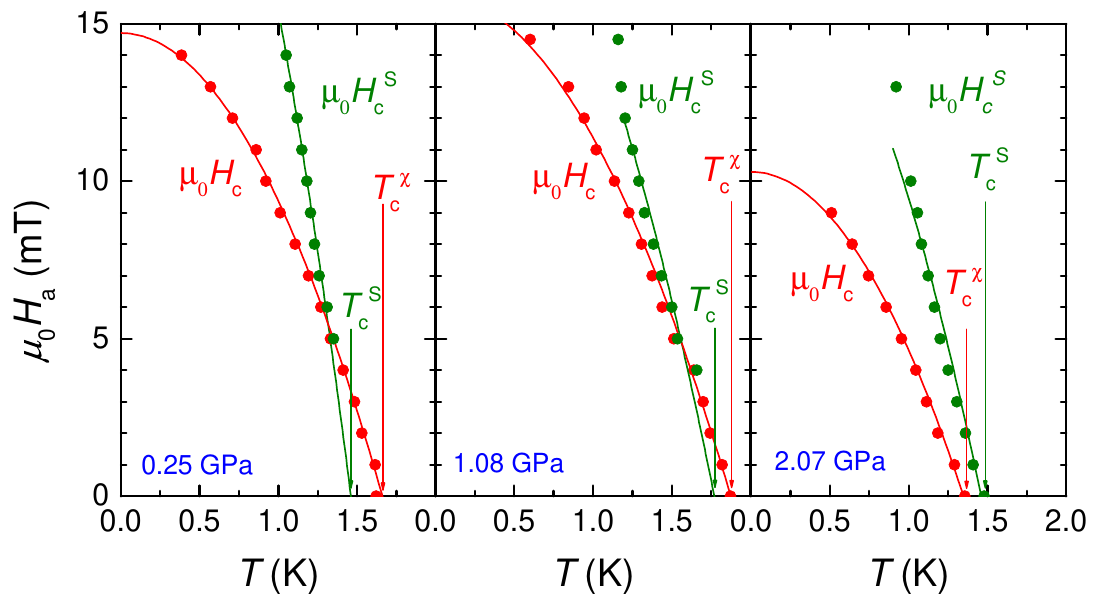}
\caption{Superconducting phase diagram of PdTe$_2$ deduced from ac-susceptibility at a pressure of 0.25~GPa (left), 1.08~GPa (middle) and 2.07~GPa (right), for $H_a$ in the basal plane. Bulk type-I superconductivity is found below the critical field $H_c (T)$. The data points (red solid symbols) follow the standard quadratic temperature variation $H_c(T) = H_c(0)[1-(T/T_c^{\chi})^2]$ (red lines). Surface superconductivity is found below $H_c^S (T)$ (green solid symbols). The transition temperature, $T_c^S$, is determined by extrapolating $H_c^S (T)$ to $H_a = 0$ (green lines). The values of the bulk $T_c^{\chi}$ and surface $T_c^S$ are indicated by arrows. Note that at the highest pressure $T_c^S > T_c^{\chi}$.}
\label{fig:figure6}
\end{figure}

Next we show how superconductivity of the surface sheath develops with pressure. Hereto we have traced $T_c^S(H)$ obtained from the $\chi_{ac}$-curves in applied fields in Fig.~\ref{fig:figure6}. Phase diagrams at 0.25, 1.08 and 2.07 GPa are presented. At 0.25~GPa we start to observe the (partial) diamagnetic screening due to the surface at a finite value $H_a \approx 5$~mT (Fig.~\ref{fig:figure4}, lower panel). The corresponding $T_c^S(H)$ points are traced in the left panel of Fig.~\ref{fig:figure6}. By extrapolating $T_c^S(H)$ to zero field we obtain $T_c^S(0)$. In the same panel we have plotted $H_c(T)$ for bulk superconductivity as well. We find $T_c^S(0) < T_c^{\chi}(0)$, just like reported previously at ambient pressure~\cite{Leng2017}. However, upon further increasing the pressure the phase lines $H_c(T)$ and $H_c^S(T)$ move apart and do no longer intersect for $p \geq 1.41$~GPa, in which case $T_c^S(0) > T_c^{\chi}(0)$. This is illustrated for $p=2.07$~GPa in the right panel of Fig.~\ref{fig:figure6}. The distinct pressure variation of $T_c^S$ and $T_c^{\chi}$ demonstrates once more that surface superconductivity is not of the standard Saint-James - de Gennes type~\cite{Saint-James&deGennes1963}. We discuss the robustness and nature of this phenomenon in the next Section.

\begin{figure}[b]
\centering
\includegraphics[width=8cm]{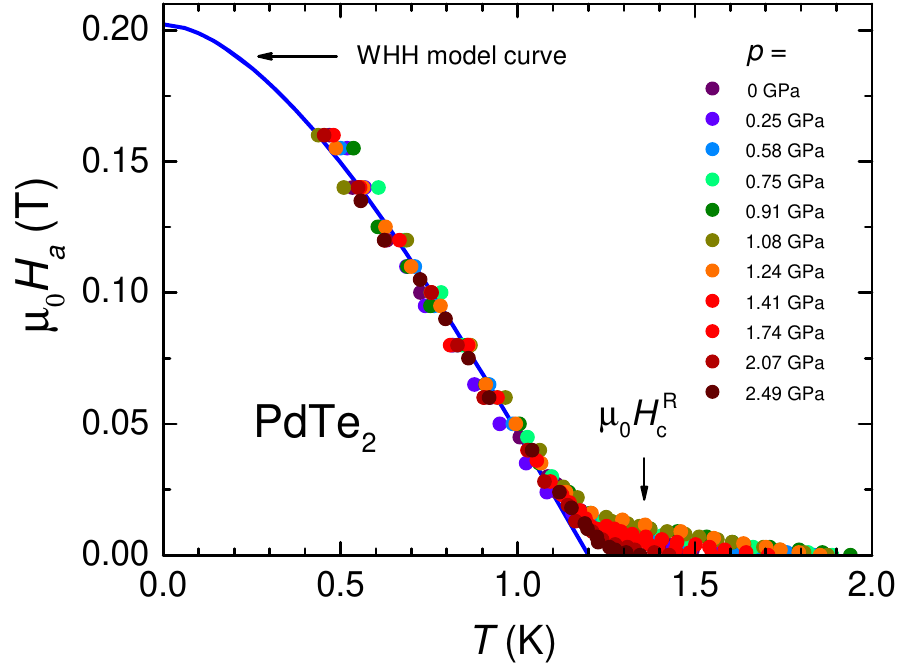}
\caption{Superconducting phase diagram of PdTe$_2$ constructed from resistance measurements in the $H$-$T$ plane at different pressures, as indicated. For 1.3-1.9~K the data points $H_c^R (T)$ denote bulk superconductivity. Below 1.3~K (partial) superconductivity persists resulting in a critical field $H_c^R (0)$ of $\simeq 0.2$~T. The blue solid line compares the data to the Werthamer-Helfand-Hohenberg model (see text).}
\label{fig:figure7}
\end{figure}

Finally we show in Fig.~\ref{fig:figure7} the $H$-$T$ phase diagrams determined from the transport data at pressures up to 2.49~GPa. At each pressure we investigated the depression of superconductivity by measuring $R(T)$ in fixed applied fields. The $R(T)$-data for 0.25~GPa are shown in the upper panel in Fig.~\ref{fig:figure4}. Additional data sets are reported in the SM~\cite{SM}. In all cases superconductivity is first depressed rapidly in small fields, and $H_c^R (T)$ tracks $H_c(T)$ for bulk superconductivity as deduced from $\chi_{ac}$ (see Fig.~\ref{fig:figure5}). The $H_c^R (T)$-data in Fig.~\ref{fig:figure7} show this behavior is restricted to the temperature range 1.3-1.9~K. Below 1.3~K the transition in $R(T)$ broadens and traces of superconductivity are visible up to $\sim 0.2$~T. By tracing in Fig.~\ref{fig:figure7} the onset temperature for superconductivity from $R(T)$ in fixed magnetic fields below 1.3~K, we observe a steady increase of $H_c^R(T)$. A comparison with the Werthamer-Helfand-Hohenberg (WHH) model~\cite{Werthamer1966} indicates the data extrapolate to $H_c^R(0) \simeq 0.2$~T for $T \rightarrow 0$. We remark that for the crystal studied in Ref.~\onlinecite{Leng2017} this value is larger, $\simeq 0.3$~T. Interestingly, $H_c^R(T)$ below 1.3~K is almost pressure independent, which shows the superconducting transition in resistance for $H_a > H_c$ is not closely connected to surface superconductivity as was proposed in Ref.~\onlinecite{Leng2017}.

\section{Analysis and Discussion}

The mechanical and electronic properties of PdTe$_2$ under pressure have been investigated theoretically by several groups~\cite{Soulard2005,Xiao2017,Lei2017}. The only experimental high-pressure study carried out so far is by Soulard \textit{et al}.~\cite{Soulard2005} who conducted high-pressure X-ray diffraction experiments at room temperature and 300~$^\circ$C to investigate the possiblity of a structural phase transition. They found that an abrupt change in the interatomic distances occurs above $p=15.7$~GPa at room temperature, but the volume \textit{versus} pressure curve exhibits no discontinuity. Under pressure the unit cell volume decreases by 17.6\% at the maximum applied pressure of 27~GPa, and the $c/a$ ratio decreases from 1.27 to 1.24 at 27~GPa. A bulk modulus, $B_0$, of 102 GPa was derived from the experimental data. This value is to be compared with 71.2~GPa (74.2~GPa) derived from first principle calculations by Lei \textit{et al}.~\cite{Lei2017} at 300~K (0~K). Xiao \textit{et al}.~\cite{Xiao2017} computed the optimized lattice parameters as a function of pressure, which are slightly overestimated compared to the experimental data~\cite{Soulard2005}. Overall, these studies indicate there is no structural transition in the modest pressure range up to 2.5~GPa in our experiments. For a layered material the change in the $c/a$-ratio is normally an important control parameter for the electronic properties. However, for PdTe$_2$ this change is very tiny and 0.2\% at most up to 2.5~GPa~\cite{Soulard2005}. In the following we focus on the superconducting properties.

\subsection{Bulk superconductivity}

A major result is the non-monotonous variation of $T_c$ with pressure reported in Fig.~\ref{fig:figure3}. $T_c$ first increases to 1.91~K at 0.91~GPa and then is gradually depressed. We first compare the experimental results with theoretical calculations. The evolution of superconductivity with pressure was investigated theoretically by Xiao \textit{et al}.~\cite{Xiao2017}. The authors used the Allen-Dynes-modified McMillan equation to calculate $T_c$, with the characteristic phonon frequency $\omega_{log}$, the electron-phonon coupling constant $\lambda$ and the Coulomb pseudopotential $\mu^* \simeq 0.1$ as input parameters. Combined electronic structure and phonon-density of states calculations show a gradual decrease of $\lambda$ and an increase of $\omega_{log}$ (blue shift), but overall the calculated $T_c$ decreases from 2.0~K at ambient pressure to 0.6~K at 10~GPa. Note the calculated $T_c$ at $p=0$ is larger than our experimental value of 1.6~K. While a decrease to 0.6~K at 10~GPa is within bounds of the extrapolation of $T_c (p)$ in Fig.~\ref{fig:figure3}, the calculations by Xiao \textit{et al}.~\cite{Xiao2017} clearly do not capture the initial increase of $T_c$ and its maximum value at 0.91~GPa. The superconducting properties of PdTe$_2$ were also investigated by Kim \textit{et al}.~\cite{Kim2018} employing the same McMillan formalism. Their phonon band structure calculations show the electron-phonon interaction is dominated by the optical $O_{1,2}$ and $O_3$ phonon modes. Furthermore, they emphasize the importance of a saddle-point van Hove singularity (vHs) close to the Fermi energy. The computed $T_c$ is 1.79~K at ambient pressure. The importance of a vHs is further illustrated by the case of PtTe$_2$, which is isoelectronic with PdTe$_2$ but does not show superconductivity. Here the vHs-band has a broad dispersion along $k_z$ leading to a lower density of states at the Fermi level and absence of superconductivity~\cite{Kim2018}. Calculations for PdTe$_2$ with a 15\% volume contraction, which corresponds to a pressure of $\sim$20~GPa, indicate the vHs band moves close to the Fermi level~\cite{Kim2018}, which would produce a higher $T_c$. However, this is at variance with the experimental data presented in Fig.~\ref{fig:figure3}.

\begin{figure}[t]
\centering
\includegraphics[width=8cm]{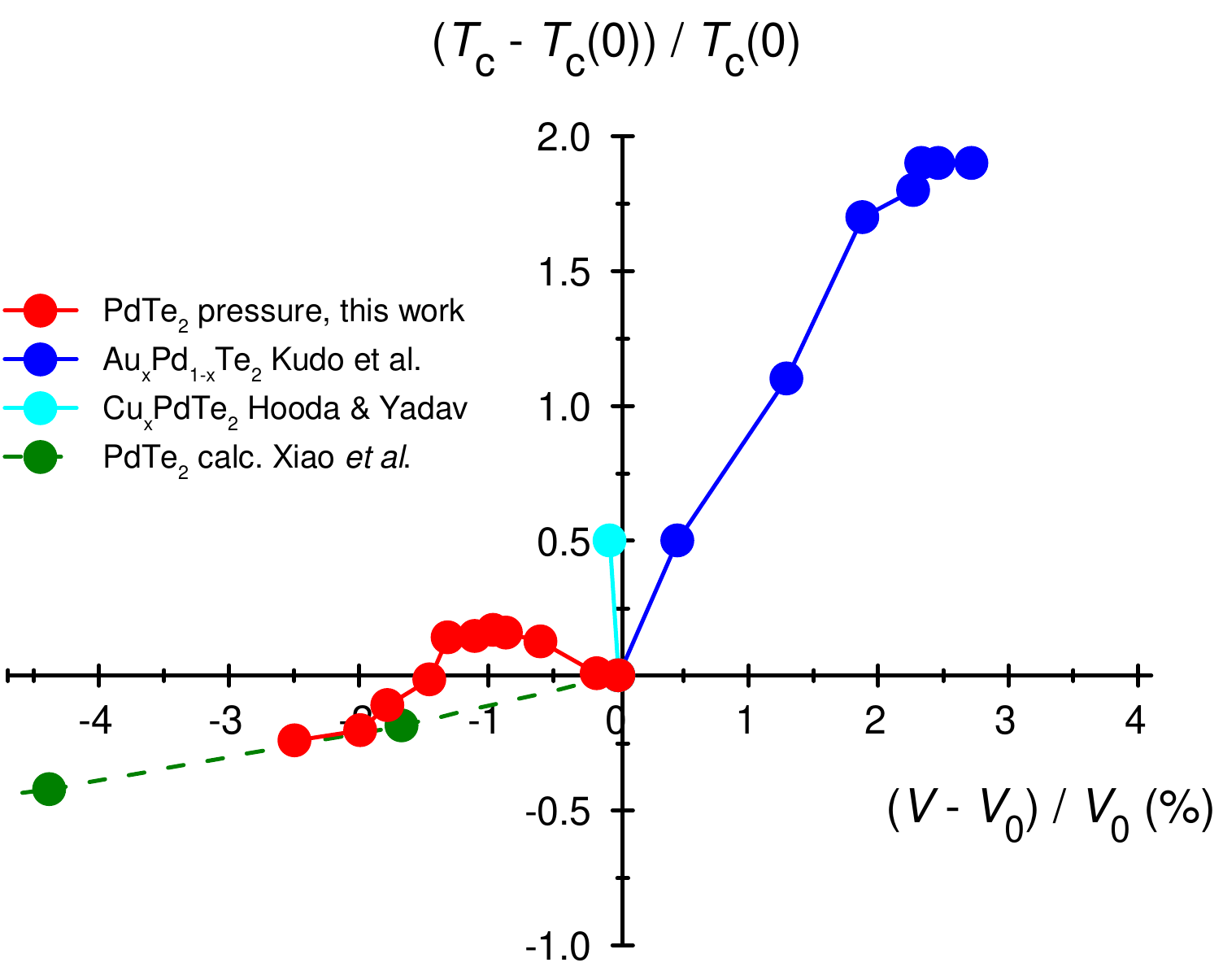}
\caption{Relative change of the superconducting transition temperature, $(T_c - T_c(0)) / T_c(0))$, as a function of the relative volume change $(V - V_0)/V_0$. Red symbols: PdTe$_2$ under pressure, this work; blue symbols: Au$_x$Pd$_{1-x}$Te$_2$, Ref.~\onlinecite{Kudo2016}; magenta symbol: Cu$_x$PdTe$_2$, Ref.~\onlinecite{Hooda&Yadav2018}; green symbols: calculated, Ref.~\onlinecite{Xiao2017}.}
\label{fig:figure8}
\end{figure}

Another way to tune $T_c$ besides pressure is via doping or substitution. Recently, it was demonstrated that Cu intercalation enhances $T_c$ to a maximum value of 2.6~K in Cu$_{x}$PdTe$_2$~\cite{Liu2015b,Ryu2015,Hooda&Yadav2018} for $x=0.06$. Upon intercalation the volume contracts, but changes are minute: $\Delta V/V = -$0.07\% for $x=0.04$~\cite{Hooda&Yadav2018}, which corresponds to an applied pressure of 0.07~GPa. This shows Cu intercalation cannot be equated to chemical pressure in tuning superconductivity. The same holds for the substitution series (Au$_{x}$Pd$_{1-x}$)Te$_2$~\cite{Kudo2016}. Upon alloying with Au, $T_c$ increases up to 4.65~K for  $x=0.40$. Simultaneously, the volume \textit{increases} by 2.5\%, which corresponds to a \textit{negative} pressure of $\sim$2.5~GPa. The experimental and calculated variation of $T_c$ with pressure and doping are summarized in Fig.~\ref{fig:figure8}. Here we trace the relative change of $T_c$ as a function of the relative volume change, $(V-V_0)/V_0$, where a bulk modulus of 102~GPa is used~\cite{Soulard2005}. Although $T_c$ generally decreases with a smaller volume, the experimentally observed positive $dT_c /dp$ for PdTe$_2$ up to 0.91~GPa is at odds with this trend.

In an attempt to shed further light on the pressure variation of $T_c$, we have conducted Hall effect measurements on two PdTe$_2$ crystals under pressure up to 2.07~GPa~\cite{SM}. At the lowest pressure of 0.25~GPa the carrier concentration, $n$, amounts to 1.5-1.7$ \times 10^{22}$~cm$^{-3}$ at 2~K. It varies quasi-linearly with pressure resulting in an increase of $\sim$20\% at 2.07~GPa. No anomalous behavior is observed around 0.9~GPa. In the most simple model the increase of $n$ is expected to result in an increase of the density of states at the Fermi level and a monotonous enhancement of $T_c$.

The non-monotonous variation of $T_c$ indicates the density of states and the electron phonon-coupling constant are affected in an intricate manner by doping and/or pressure. Possibly this is a result from band structure subtleties that have not been probed in the coarse-grained calculations carried out so far~\cite{Soulard2005,Xiao2017,Lei2017}. In order to access the electronic band structure under pressure, a quantum oscillations study is highly desirable. The feasibility to observe the Shubnikov - de Haas effect and the de Haas - van Alphen effect at ambient pressure has been demonstrated in Refs.~\onlinecite{Dunsworth1975,Fei2017,Zheng2018}. In the same context, small structural modifications that might influence $T_c$, such as changes in the $z$-coordinate of Te atoms in the unit cell that would affect the $O_{1,2}$ and $O_3$ phonon modes, cannot be excluded based on the X-ray diffraction experiment with a first pressure point at 2.2~GPa~\cite{Soulard2005}. This calls for high-precision low-pressure ($p \leq 2.5$~GPa) single-crystal X-ray diffraction measurements.

\subsection{Surface superconductivity}

The distinct pressure variation of the superconducting transition temperature of the surface sheath, $T_c^S$, and of the bulk, $T_c^{\chi}$, reported in Figs.~\ref{fig:figure3} and ~\ref{fig:figure6}, is an extraordinary result. We recall this feature is derived from the ac-susceptibility curves measured in fixed magnetic fields at eleven different pressures. Selected data sets at 0.25~GPa are shown in Fig.~\ref{fig:figure4} and at 1.08 and 2.07~GPa in the SM~\cite{SM}. The data show how type-I superconductivity in the bulk, probed by the DPE-peaks in small applied dc-fields, is progressively depressed with field, while surface superconductivity is observed for $H_a > H_c$ (see also Ref.~\onlinecite{Leng2017}). Upon increasing the pressure, the DPE peak is more rapidly depressed compared to surface screening. At 2.07~GPa the DPE effect is - already in the lowest applied fields - almost completely screened by the surface~\cite{SM}. Hence for $p \geq 1.41$~GPa $T_c^S > T_c^{\chi}$. This is further underpinned by the observation that $H_c(T)$, defined by $T_c^{\chi}(H)$, follows the quadratic temperature variation at all pressures, characteristic for bulk type-I superconductivity (Fig.~\ref{fig:figure5}). Note that $T_c^{S}$ is defined as the onset temperature for  the diamagnetic signal due to surface superconductivity, while the transition itself may become very broad. $H_c^S(p)$ has a maximum near 0.9~GPa, similar to $H_c (p)$, as reported in the SM~\cite{SM}. When the $H_c^S(T,p)$ data is traced in the reduced form $h^*(t)$ the data do not collapse on a single curve as, see SM~\cite{SM}. Instead the trend is that the values $h^*(t)$ increase with respect to pressure, which indicates the superconducting pairing interaction changes in a non-trivial way. The distinct $H_c(T)$- and $H_c^S$-curves and their dissimilar pressure dependence strongly suggest surface and bulk superconductivity are independent phenomena and not tightly connected, in contrast to the familiar Saint James - de Gennes surface superconductivity~\cite{Saint-James&deGennes1963}. It remains tempting to relate surface superconductivity in PdTe$_2$ to topological surface states detected by ARPES\cite{Liu2015a,Noh2017,Clark2017}. These surface states could possibly be investigated by STM experiments in small applied fields ($H_a > H_c$). The STM experiments performed so far were predominantly directed to probe bulk superconductivity~\cite{Das2018,Clark2017}. Moreover, for the spectra taken in a magnetic field the intermediate state that occurs below $H_c$ for a finite demagnetization factor was not taken into account.

In the resistance measurements (partial) superconductivity is observed up to about 0.2~T for $T \rightarrow 0$ (Fig.~\ref{fig:figure7}), a value that largely exceeds $H_c(0)$ and $H_c^S(0)$. The enhanced $H_c^R(T)$-curves below 1.3~K are quasi pressure independent. By extrapolating the data in this field range to $H_a \rightarrow 0$ with the WHH function a pressure independent $T_c = 1.2$~K is found. Since $T_c^S$ has a pronounced pressure variation the resistive superconducting transitions measured in this field range are not connected to surface superconductivity. Note that for the crystal studied in Ref.~\onlinecite{Leng2017} it was concluded that the transport experiment does probe surface superconductivity, but these experiments were performed at ambient pressure only. The persistence of superconductivity in resistance measurements in field is puzzling. Normally such an effect is attributed to filamentary superconductivity. Its pressure independence indicates it might not be intrinsic to PdTe$_2$.

\section{Summary and conclusions}

We have carried out a high-pressure transport and ac-susceptibility study of superconductivity in the type-I superconductor PdTe$_2$ ($T_c = 1.64$~K). $T_c$ shows a pronounced variation with pressure: it increases at low pressure, then passes through a maximum of 1.91~K around 0.91~GPa, and subsequently decreases smoothly up to the highest pressure measured, $p_{max} = 2.5$~GPa. The critical field, $H_c$, follows a similar behavior, leading the $H_c(T)$-curves at different pressures to collapse on a single universal curve with the characteristic quadratic in temperature depression of $H_c$ for type-I superconductivity. Type-I superconductivity is robust under pressure. In view of the absence of structural modifications in our pressure range and the minute change of the $c/a$-ratio ~\cite{Soulard2005}, the non-monotonous variation of $T_c$ indicates an intricate role of the dominant phonon frequency, the electron-phonon-coupling parameter and Coulomb pseudopotential used to compute $T_c$ with help of the McMillan formula. This effect has not been captured by band structure calculations so far~\cite{Xiao2017,Kim2018}, notably the electron band structure calculations predict a smooth decrease of $T_c$ under pressure~\cite{Xiao2017}. This calls for more elaborate and detailed calculations for pressures up to $p_{max} = 2.5$~GPa.

The unusual surface superconductivity, first reported at ambient pressure~\cite{Leng2017}, persists under pressure. Surprisingly, for $p \geq 1.41$~GPa the superconducting transition temperature for the surface $T_c^S$ exceeds $T_c$ of the bulk. This tells us surface and bulk superconductivity are distinct phenomena. This is further confirmed by the observation that the phase lines $H_c(T)$ and $H_c^S(T)$ move apart under pressure and no longer intersect for $p \geq 1.41$~GPa. We propose surface superconductivity possibly has a non-trivial nature and originates from topological surface states detected by ARPES\cite{Liu2015a,Noh2017,Clark2017}. This calls for quantum-oscillation experiments under pressure, possibly enabling one to follow the pressure evolution of the bulk electronic structure and topological surface states.

In the same spirit it will be highly interesting to extend the experiments to higher pressures, especially because a pronounced change in the electronic properties of PdTe$_2$ is predicted to occur in the range  4.7-6.1~GPa: the type-II Dirac points disappear at 6.1~GPa, and a new pair of type-I Dirac points emerges at 4.7~GPa~\cite{Xiao2017}. Thus a topological phase transition may occur in the pressure range 4.7-6.1~GPa. This in turn might have a strong effect on (surface) superconductivity, because the tilt of the Dirac cone vanishes~\cite{Fei2017,Shapiro2018}. We conclude further high-pressure experiments on PdTe$_2$ provide a unique opportunity to investigate the connection between topological quantum states and superconductivy.

\vspace{6mm}

Acknowledgements: H.L. acknowledges the Chinese Scholarship Council for Grant No.~201604910855. This work was part of the research program on Topological Insulators funded by FOM (Dutch Foundation for Fundamental Research on Matter). It was further supported by the JSPS (Japan Society for the Promotion of Science) Program for Fostering Globally Talented Researchers, Grant Number R2903.

\bibliography{References_PdTe2}

\bibliographystyle{apsrev4-1}
%\end{multicols}{2}

\clearpage
%\begin{multicols}{1}

\noindent
{\Large \textbf{Supplemental Material for ``Superconductivity under pressure in the Dirac semimetal PdTe$_2$''}}
\vspace{10mm}

%\author{H. Leng} \affiliation{Van der Waals - Zeeman Institute, University of Amsterdam, Science Park 904, 1098 XH Amsterdam, The Netherlands}
%\author{A. Ohmura} \affiliation{Pacific Rim Solar Fuel System Research Center, Niigata University, 8050, Ikarashi 2-no-cho, Nishi-ku, Niigata, 950-2181, Japan}
%\affiliation{Faculty of Science, Niigata University, 8050, Ikarashi 2-no-cho, Nishi-ku, Niigata, 950-2181, Japan}
%\author{L. N. Anh} \affiliation{International Training Institute for Materials Science, Hanoi University of Science and Technology, 1 Dai Co Viet Road, Ha Noi, Vietnam}
%\author{F. Ishikawa} \affiliation{Faculty of Science, Niigata University, 8050, Ikarashi 2-no-cho, Nishi-ku, Niigata, 950-2181, Japan}
%\author{T. Naka} \affiliation{National Institute for Materials Science, Sengen 1-2-1, Tsukuba, Ibaraki 305-0047, Japan}
%\author{Y. K. Huang} \affiliation{Van der Waals - Zeeman Institute, University of Amsterdam, Science Park 904, 1098 XH Amsterdam, The Netherlands}
%\author{A. de Visser} \affiliation{Van der Waals - Zeeman Institute, University of Amsterdam, Science Park 904, 1098 XH Amsterdam, The Netherlands}

%\date{\today}

%\maketitle

\newcommand{\beginsupplement}{%
        \setcounter{table}{0}
        \renewcommand{\thetable}{S\arabic{table}}%
        \setcounter{figure}{0}
        \renewcommand{\thefigure}{S\arabic{figure}}%
    }
\beginsupplement
\vspace{10mm}
\noindent
\textmd{\textbf{1. Resistance and ac-susceptibility measurements in field}}
\noindent

In order to investigate the response of the superconducing phase of PdTe$_2$ to an applied magnetic field and to construct the field-temperature phase diagram electrical resistivity and ac-susceptibility measurements were carried out. The resistance as a function of temperature, $R(T)$, was measured using a sensitive resistance bridge (model Linear Research LR700) in a four-point geometry by a low-frequency ac-method with an excitation current $I= 300~\mu$A. The ac-susceptibility was measured by placing the crystal in a small coil-set with an excitation and pick-up coil, mounted inside the pressure cell. The excitation field was $\mu_0 H_{ac} = 0.0047$~mT. The in-phase and out-of-phase signals were recorded at a frequency of $f_{ac} = 313$~Hz using a lock-in amplifier (EG\&G Instruments Model 7260). The applied dc-field, directed in the basal plane of the crystal, was produced by a superconducting magnet. Special care was taken to reduce the remnant field of the superconducting magnet to close to zero, since the PdTe$_2$ crystals show type-I superconductivity. Measurements were carried out at eleven different pressures. The data at 0.25~GPa are reported in the main text. In Fig.~S1 and Fig.~S2 selected data sets at 1.08 and 2.07~GPa, respectively, are presented.

\begin{figure}  
\centering
\includegraphics[width=8.5cm]{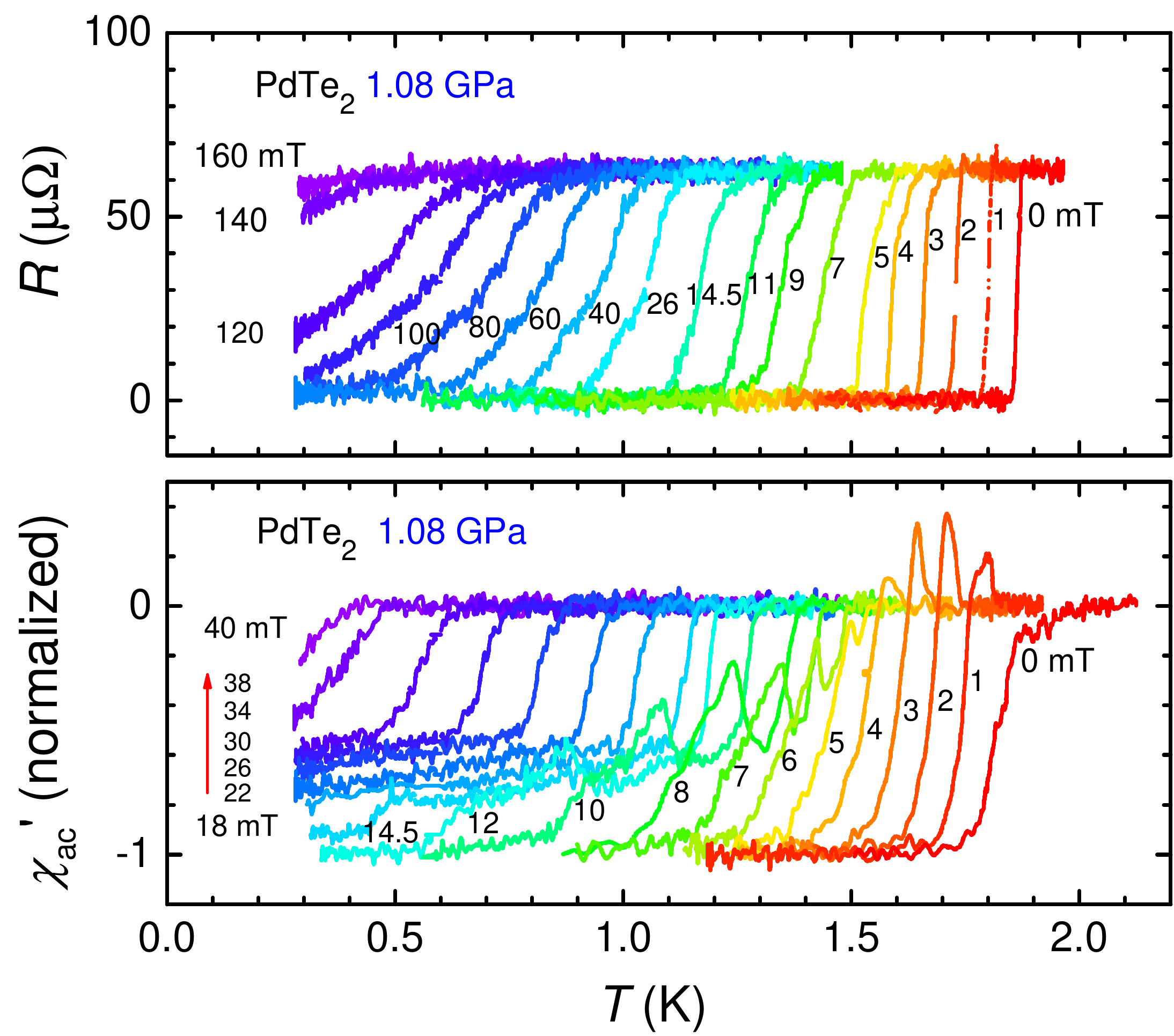}
\caption{Upper panel: Resistance of PdTe$_2$ as a function of temperature at a pressure $p = 1.08$~GPa measured in applied magnetic fields $\mu_0 H_a \parallel I \parallel a$. Curves from right to left are taken in fields of 0, 1, 2, 3, 4, 5, 7, 9, 11, 14.5, 26, 40, 60, 80, 100, 120, 140 and 160~mT. Lower panel: Ac-susceptibility at $p = 1.08$~GPa measured in applied magnetic fields. Curves from right to left in 0~mT to 8~mT with 1~mT steps and in 10, 12, 14.5, 18, 22, 26, 30, 34, 38 and 40~mT. }
\end{figure}

\begin{figure}  
\centering
\includegraphics[width=8.5cm]{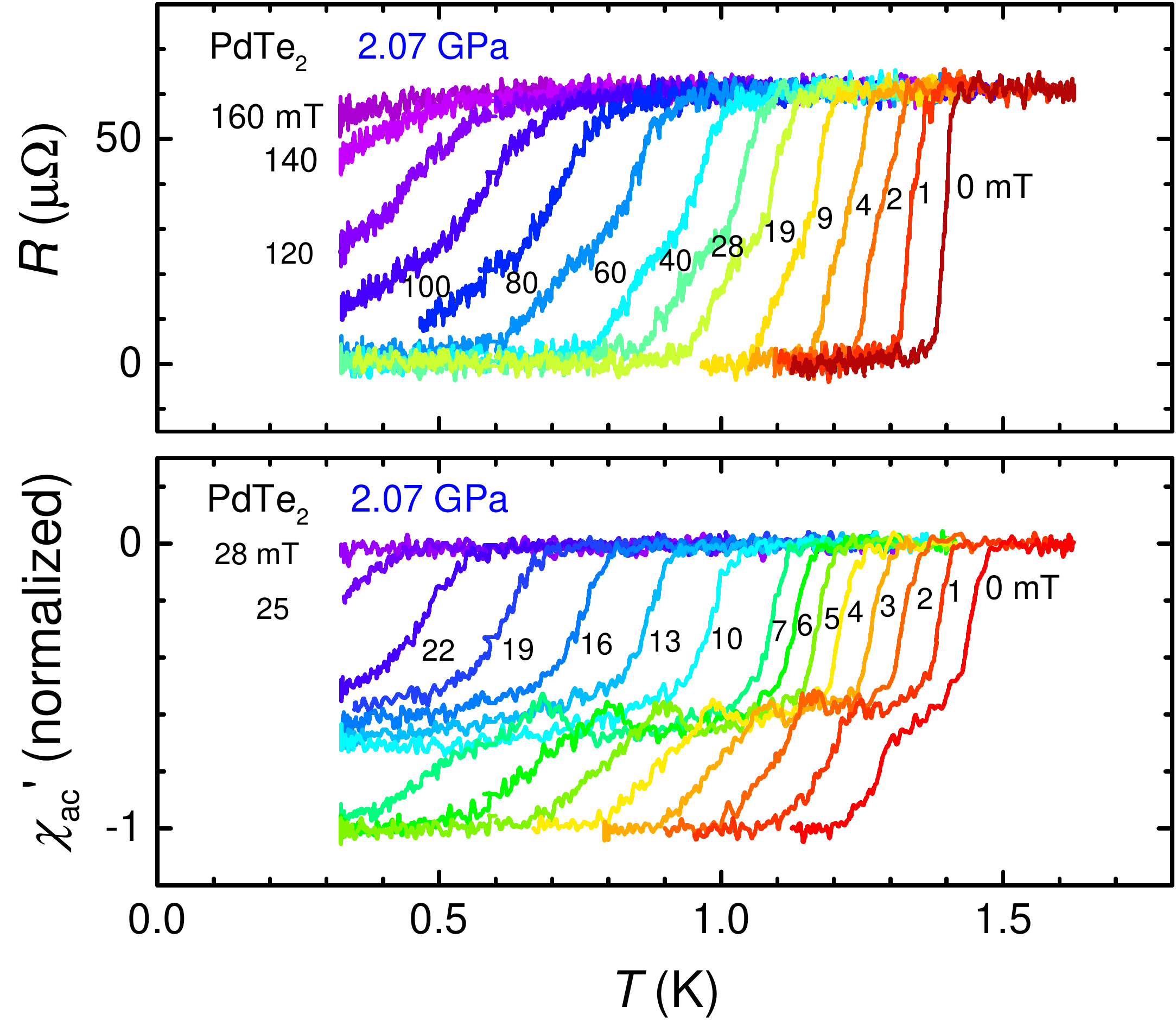}
\caption{Upper panel: Resistance of PdTe$_2$ as a function of temperature at a pressure $p = 2.07$~GPa measured in applied magnetic fields $\mu_0 H_a \parallel I \parallel a$. Curves from right to left are taken in fields of 0, 1, 2, 4, 9, 19, 28, 40, 60, 80, 100, 120, 140 and 160~mT. Lower panel: Ac-susceptibility at $p = 2.07$~GPa measured in applied magnetic fields. Curves from right to left in 0~mT to 7~mT with 1~mT steps and in 10, 13, 16, 19, 22, 25 and 28~mT.}
\end{figure}

We first discuss the data at 1.08~GPa. The transition to $R=0$ in small fields is sharp and rapidly depressed up to $\sim 11$~mT. This signals bulk type-I superconductivity, and by extrapolating the data in the temperature range 1.3-1.9~K using a quadratic temperature variation we estimate a critical field $H_c^R (0) = 22$~mT (see Fig.~7 in the main text and Fig.~S4). For larger fields superconductivity is depressed at a much lower rate, the transition broadens and becomes incomplete for $H_a \geq 120$~mT. Signs of superconductivity in $R(T)$ persist up to 200~mT for $T \rightarrow 0$. $H_c^R(T)$ extracted from the upper panel in Fig.~S1 is reported in Fig.~7 in the main text. The $\chi_{ac}$-data, plotted in the lower panel of Fig.~S1, shows the peak due to the differential paramagnetic effect (DPE) is rapidly depressed with field. The DPE peak is due to the intermediate state in the bulk of the type-I superconductor. From the shift of the DPE peak we determine the $H_c$ phase boundary. However, for $H_a > H_c$ large screening signals persist. This we attribute to superconductivity of the surface sheath~\cite{Leng2017} with a critical field $H_c^S$. A sizeable screening is still observed at 40~mT. The phase boundaries $H_c$ and $H_c^S$ derived from $\chi_{ac}$ are reported in Fig.~6 (middle panel) in the main text.

The resistance data at 2.07~GPa, shown in the upper panel of Fig.~S2, compare well to the data at 1.08~GPa, except superconductivity in zero field is further depressed from $T_c^R =1.85$~K at 1.08~GPa to 1.41~K at 2.07~GPa. Consequently, the extrapolated critical field for bulk superconductivity is reduced to $H_c^R (0) = 19$~mT. The $\chi_{ac}$-data plotted in the lower panel, however, show a very different behavior compared to the data at 1.08~GPa. The DPE peak is reduced at 2.07~GPa and appears in the data in applied dc-fields well below the initial diamagnetic step. This we attribute to the notion that surface screening precedes screening due to bulk superconductivity. Thus $T_c^S > T_c^{\chi}$, where $T_c^{\chi}$ is the bulk superconducting transition temperature. The data points extracted from Fig.~S2 in this manner define the phase boundaries $H_c (T)$ and $H_c^S (T)$ reported in Fig.~6 (right panel) in the main text. Screening at the surface is not complete and amounts to 60\% only. Note the DPE peak is no longer observed for $H_a > 10~$mT, and $H_c (T)$ (defined by $T_c^{\chi}(H)$) follows the quadratic temperature variation for the bulk type-I superconducting phase.

\vspace{10mm}
\noindent
\textmd{\textbf{2. Critical field of surface superconductivity}}
\noindent

At each pressure we have constructed the $H_c^S (T)$ phase boundary. In an attempt to collapse all the $H_c^S(T,p)$-data on a single curve a plot of $h^*(t)$ is presented in Fig.~S3, where $h^* =(H_c^S(T)/T_c^S)/(-dH_c^S/dT)|_{T_c^S}$ and $t=T/T_c^S$. Note that for pressures up to 1.24~GPa $T_c^S$ and the initial slope $-dH_c^S/dT|_{T_c^S}$ are determined by extrapolation, as shown in Fig.~6 (main text) for 0.25 and 1.08~GPa. This introduces some uncertainty in the data, but the overall trend is that $h^*(0)$ increases with respect to pressure. This indicates the superconducting pairing interaction changes in a non-trivial way. Leng \textit{et al}.~\cite{Leng2017} reported that the $H_c^S (T)$-curve at ambient pressure follows a quadratic temperature variation. Such a behavior is absent for the present crystal. Instead $H_c^S (T)$ rather shows a downward or upward curvature near $t=0.7-0.8$.

\begin{figure} [h]
\centering
\includegraphics[width=8cm]{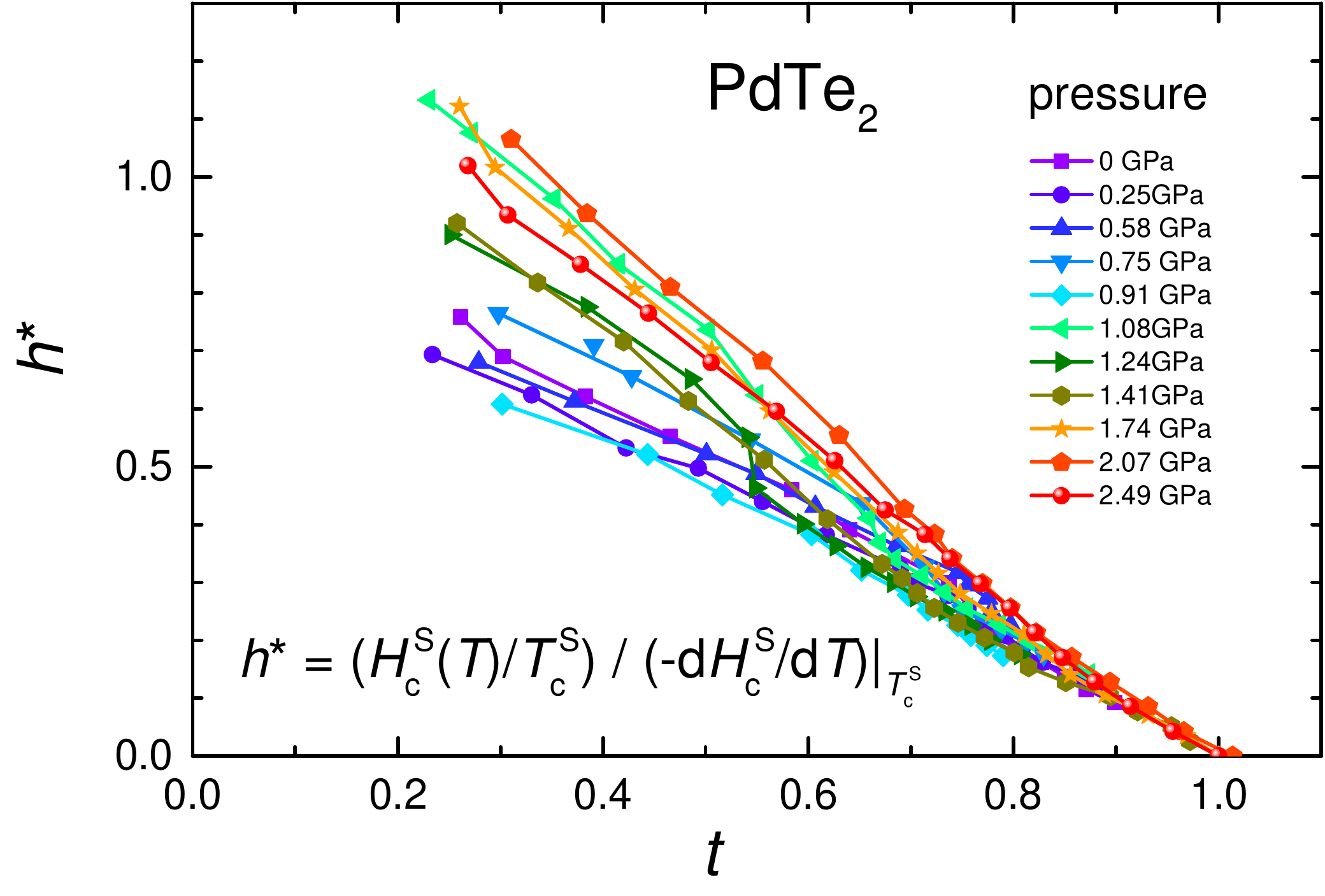}
\caption{Critical field $H_c^S(T)$ for superconductivity of the surface sheath in PdTe$_2$ at pressures between 0 and 2.49~GPa as indicated. The data are plotted in the reduced form $h^* =(H_c^S(T)/T_c^S)/(-dH_c^S/dT)|_{T_c^S}$ \textit{versus} $t=T/T_c^S$. }
\end{figure}

\noindent
\vspace{10mm}
\textmd{\textbf{3. Pressure variation of the critical field}}
\noindent

In Fig.~S4 the pressure variation of the critical fields $H_c$ and $H_c^R$ in the limit $T \rightarrow 0$ and at $T=0.3$~K for $H_c^S$ is presented. The field $H_c (0)$ is representative of bulk superconductivity. It is determined from ac-susceptibility with help of the expression $H_c(T) = H_c(0)[1-(T/T_c)^2]$, where $T_c = T_c^{\chi}$. The $H_c(T)$-curves measured at eleven different pressures are reported in Fig.~5 of the main text. $H_c^R(0)$ is determined from the data in the temperature range 1.3-1.9~K in Fig.~7 (main text) by extrapolating $T \rightarrow 0$, using the quadratic temperature variation with $T_c = T_c^R$. Note the temperature range in which $H_c^R(T)$ represents type-I superconductivity and follows a quadratic temperature variation is small, since below $T = 1.3$~K $H_c^R(T)$ shows a pronounced upturn (see Fig.~7 in the main text). Consequently, the fit brings about an uncertainty in $H_c^R(0)$, which explains the overestimated values compared to $H_c(0)$. $H_c^S(0)$ represents the critical field at $T=0.3$~K for superconductivity of the surface sheath determined by ac-susceptibility. For all three data sets in Fig.~S4 a maximum in the critical field as a function of pressure is observed near 0.9-1.2~GPa.

\begin{figure}  
\centering
\includegraphics[width=7cm]{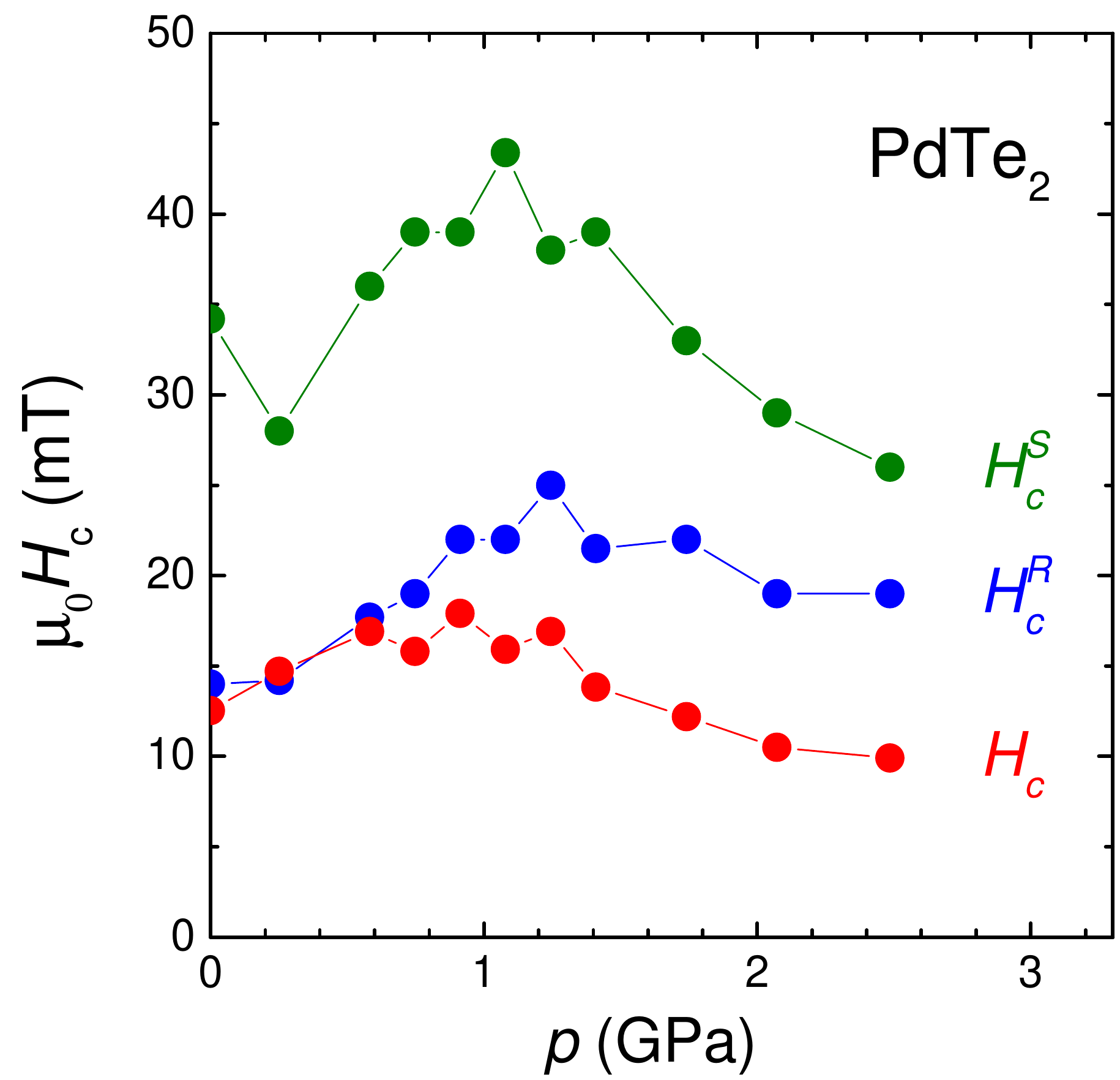}
\caption{Pressure variation of the critical field of PdTe$_2$. $H_c$ (red symbols) represents bulk type-I superconductivity determined by $\chi_{ac}$ measurements in the limit $T \rightarrow 0$. $H_c^R $ (blue symbols) is determined from resistance measurements by extrapolating the initial low field $H_c^R(T)$-data to 0~K using a quadratic temperature variation. $H_c^S$ (red symbols) represents surface superconductivity at the lowest temperature, $T=0.3$~K, as extracted from $\chi_{ac}$ .}
\end{figure}

\vspace{10mm}
\noindent
\textmd{\textbf{4. Hall-effect measurements}}
\noindent

\begin{figure} 
\centering
\includegraphics[width=8cm]{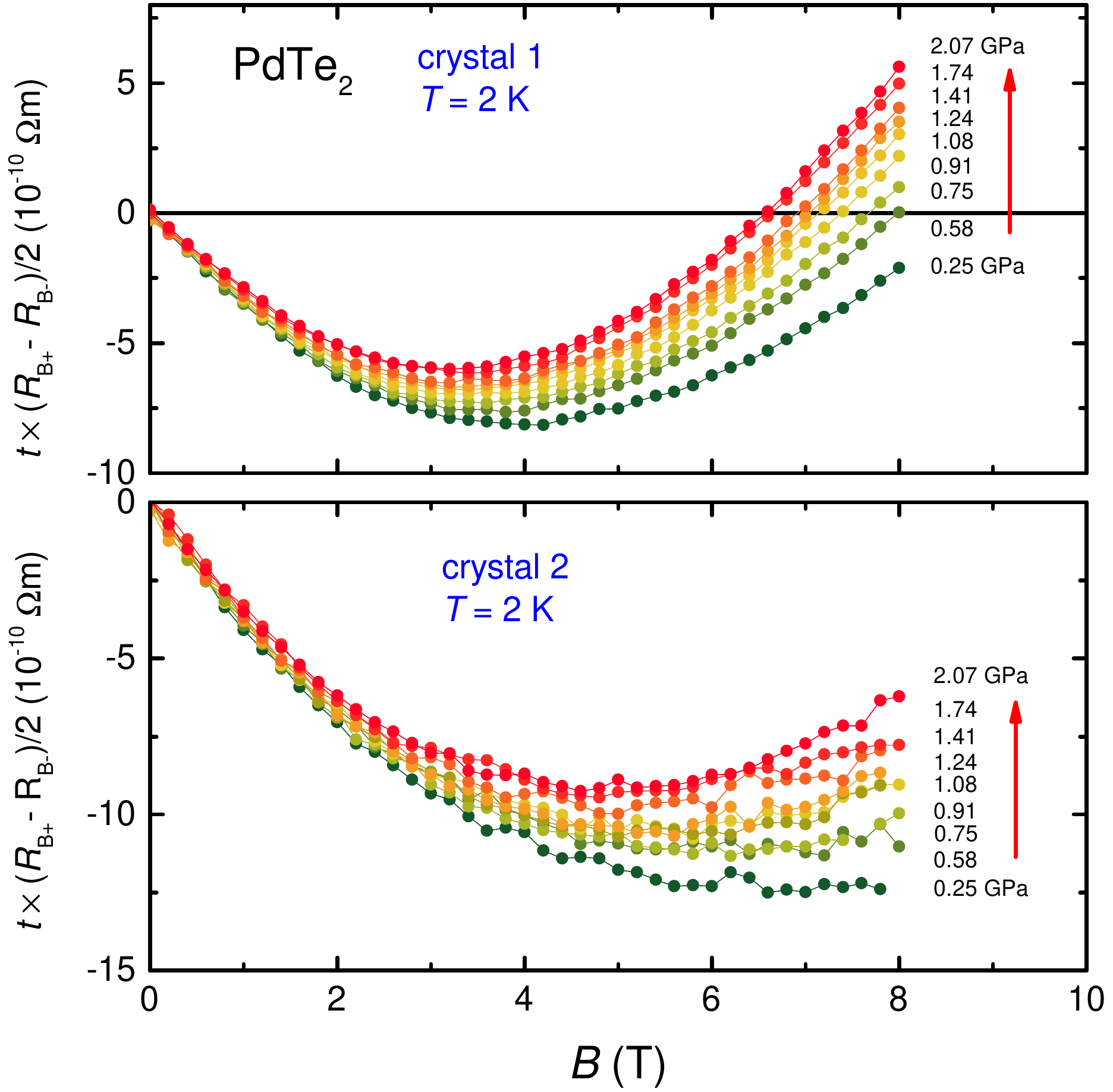}
\caption{Symmetrized Hall resistance multiplied by the crystal thickness $t$ as a function of magnetic field for two PdTe$_2$ crystals at pressures ranging from 0.25 to 2.07~GPa, as indicated. The temperature is 2~K.}
\end{figure}

\begin{figure} [b]
\centering
\includegraphics[width=8cm]{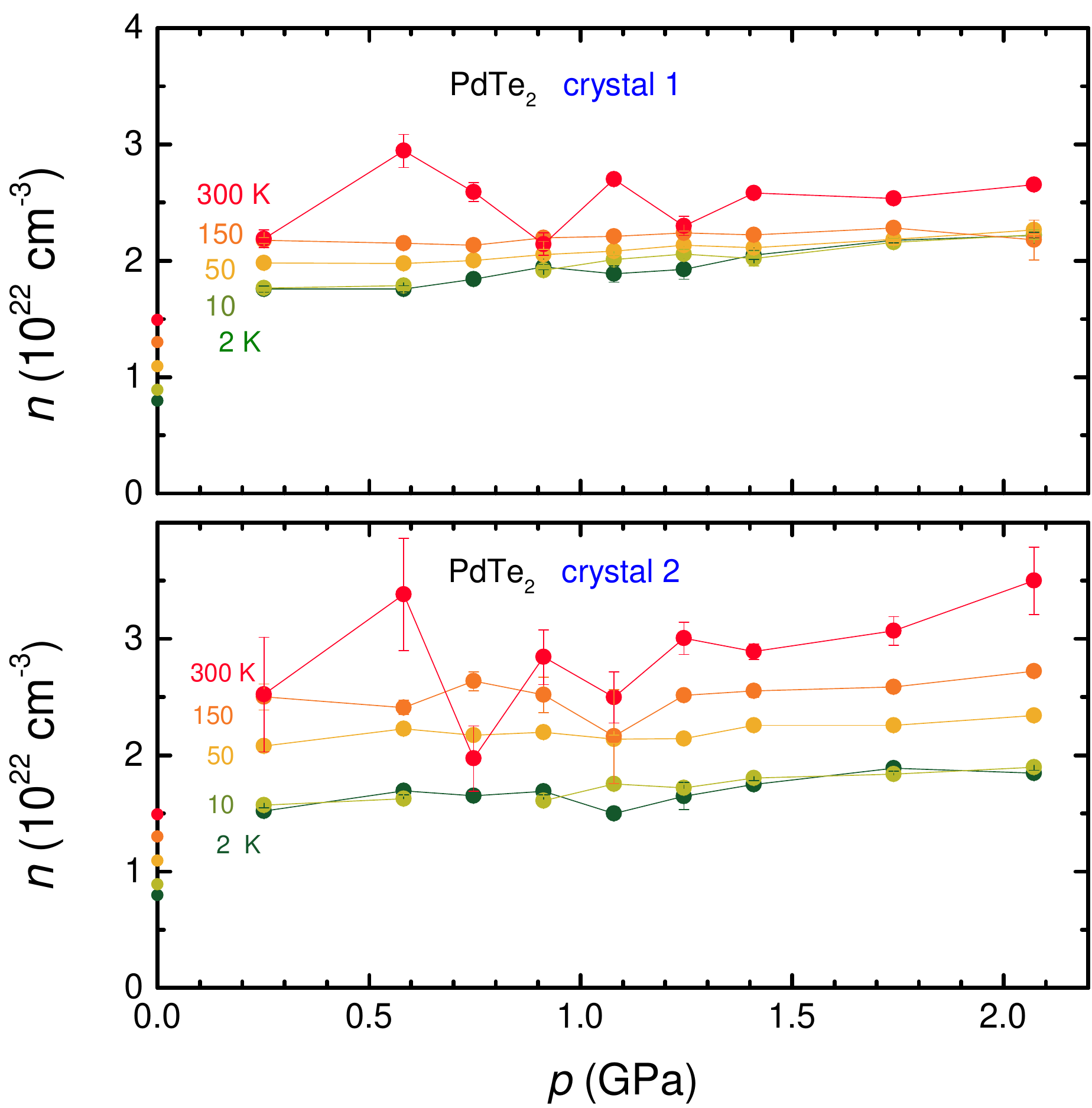}
\caption{Carrier concentration as a function of pressure for two PdTe$_2$ crystals, at temperatures of 2, 10, 50, 150 and 300~K, as indicated. The data at ambient pressure are taken from a third crystal.}
\end{figure}

The Hall effect was measured on two PdTe$_2$ crystals in a piston-cylinder clamp cell developed for the Physical Property Measurement System (PPMS, Quantum Design) at nine different pressures up to 2.07~GPa. The sample space is 4.4~mm in diameter and $\sim$15~mm in height. Two crystals were placed in two stages along a compression axis perpendicular to the sample plane. The sample size (length $\times$ width $\times$ thickness) amounts to $2.8 \times 1.4 \times 0.08$~mm$^3$ and $2.9 \times 1.0 \times 0.19$~mm$^3$ for crystal 1 and 2, respectively. The current was applied in the basal-plane of the crystals, whereas the magnetic field was applied along the trigonal axis, perpendicular to the sample plane. Data were collected at temperatures of 2, 10, 50, 150 and 300~K in magnetic fields up to 8~T. Measuremenst were carried out for two field polarities, $B^+$ and $B^-$, and the Hall resistance, $R_H$, was obtained by symmetrizing:  $R_H = (R_{B^+} - R_{B^-})/2$. In Fig.~S5 we show $t \times R_H$ as a function of the applied field at 2~K at different pressures. Here $t$ is the sample thickness. $R_H (B)$ is a non-linear function indicating the presence of several charge carrier bands, expected from Fermi surface measurements~\cite{Dunsworth1975,Zheng2018}. For crystal~1 the Hall resistance goes through a deep minimum and changes sign in the field range 6-8~T. For crystal~2 the minimum is less pronounced. We estimate the carrier concentration, $n$, from the initial linear slope of $R_H (B)$. The results are traced in Fig.~S6. Upon lowering the temperature from 300~K to 2~K, $n$ drops typically by 20\% and 50\% for crystal 1 and 2, respectively. At 2~K, $n$ amounts to 1.5-1.7$ \times 10^{22}$~cm$^{-3}$ at 0.25~GPa. It varies quasi-linearly with  pressure and has increased by $\sim$20\% at the highest pressure. No anomalous behavior is observed around 0.9~GPa, where $T_c(p)$ has a maximum. We also measured the Hall resistance at ambient pressure on a third crystal. The resulting carrier concentration is 0.8$ \times 10^{22}$~cm$^{-3}$ at 2~K. which is about a factor two smaller compared to the values for crystal 1 and 2.

%\end{multicols*}{1}

%\bibliography{References_PdTe2}

%\bibliographystyle{apsrev4-1}

\end{document}